\documentclass[preprint,aps]{revtex4}
\begin{document}
\title{SUSY SHAPE-INVARIANT HAMILTONIANS FOR THE GENERALIZED
DIRAC-COULOMB PROBLEM}
\author{
R. de Lima Rodrigues\thanks{e-mail: rafaelr@cbpf.br or
rafael@df.ufcg.edu.br.}\\
Unidade Acad\^emica de Educa\c{c}\~ao,
Universidade Federal de Campina Grande\\
Cuit\'e - PB, 58.175-000 - Brazil \\
Centro Brasileiro de Pesquisas F\'\i sicas (CBPF)\\
Rua Dr. Xavier Sigaud, 150, CEP 22290-180, Rio de Janeiro, RJ, Brazil\\
Arvind Narayan Vaidya (In memory)\\
Instituto de F\'\i sica - Universidade Federal do Rio de Janeiro \\
Caixa Postal 68528 - CEP 21945-970, Rio de
Janeiro, Brazil}

\begin{abstract}
A spin $\frac 12$ relativistic particle described by a general
potential in terms of the sum of the Coulomb potential with a
Lorentz scalar potential is investigated via supersymmetry in
quantum mechanics.

\vspace{1cm}


\vspace{3cm} ${}^{\beta}$  e-mail: rafaelr@cbpf.br or
rafael@df.ufcg.edu.br. This work was presented at XIV EVJAS, section
particles and fields, from 21/january to 02/February (2007), in
Campos do Jord\~ao-SP, Brazil. Preprint CBPF-NF-028/08, www.cbpf.br.
\end{abstract}

\maketitle

\newpage

Supersymmetry in Quantum Mechanics (SUSY QM) \cite{W81} is of
intrinsic mathematical interest in its own as it
connects otherwise apparently unrelated second-order differential equations.

The (1+3) and (1+1) dimensional Dirac equations with both
scalar-like and vector-like potentials are well known in the
literature for a long time \cite{Zhong85}.
The connection between position-dependent-effective-mass and shape
invariant condition under parameter translation
 has been discussed in non-relativistic quantum
mechanics \cite{quesne04,bagchi05}. Recently, some relativistic
shape invariant potentials have been investigated \cite{Inv01}.

Exact solutions for the bound states in this mixed potential can be
obtained by the method of separation of variables
\cite{gre90,Tutik92,Ik93} and also by the use of the dynamical
algebra $SO(2,1)$ \cite{pan95}. In a recent paper the solution of
the scattering problem for this potential has been obtained by an
analytic method and also by an algebraic method \cite{AL02}, the
problem of a relativistic Dirac electron with a $1/r$ scalar
potential, as well as a Dirac magnetic monopole and an Aharonov-Bohm
potential has also been investigated \cite{villalba05}, and the
bound eigenfunctions and spectra of a Dirac hydrogen atom have been
found via $su(1,1)$ Lie algebra \cite{romero05}.

 Recently exact solutions have been found for fermions in the presence
of a classical background which is a mixing of the time-dependent of
a gauge potential and a scalar potential \cite{chen05}. Also,
exactly solvable Eckart scalar and vector potentials in the Dirac
equation have been investigated via SUSY QM \cite{jia05}, the
$S$-wave Dirac equation has been solved exactly for a single
particle with spin and pseudospin symmetry moving in a central
Woods-Saxon potential \cite{guo05}.

 The
special case of the non-relativistic \cite{elso} and relativistic
Coulomb problems have been treated recently via SUSY QM
\cite{R04,Tamar-05,Hosho06}. In this work, the
relativistic Coulomb potential with a Lorentz scalar potential is
investigated via shape invariance conditions of the SUSY QM.

The time independent Dirac equation may be written in the form
$
H\Psi= E\psi,
$where the Hamiltonian is given by
$$
\label{h1} H= \rho_1\otimes\vec\sigma\cdot\vec p +
\left(M-\frac{A_2}{r} \right)\rho_3\otimes{\bf
1}_{2\hbox{x}2}-\frac{A_1}{r}\otimes\ {\bf 1}_{4\hbox{x}4},
$$
and we have used a direct product notation in which $\rho_i$ and
$\sigma_i, (i= 1,2,3)$ are the Pauli spin matrices obeying $[\rho_i,
\sigma_j]_-= 0,$ with $\hbar=c=1.$

We consider \cite{BD65}

\begin{equation}
\Psi= \left(\begin{array}{cc}
\frac{iG_{\ell j}}{r}\phi^{\ell}_{jm}\\
\frac{F_{\ell j}}{r}
\vec\sigma\cdot\vec  n\phi^{\ell}_{jm}
\end{array}\right),
\end{equation}
where $\phi^{\ell}_{jm}=\phi^{(\pm)}_{jm},$ for $j= \ell\pm\frac
12.$ Next, using the relation $[{\bf 1}+\vec\sigma\cdot\vec L,
\vec\sigma\cdot\vec n]_+= 0 $ we obtain $K\Psi= -k\Psi$ and the
following radial equations

\begin{eqnarray}
\frac{dG_{\ell j}}{dr}+\frac{k}{r}G_{\ell j}-
\left(E+M-\frac{A_2}{r}+\frac{A_1}{r}\right)F_{\ell j}&&= 0,\nonumber\\
\frac{dF_{\ell j}}{dr}-\frac{k}{r}F_{\ell j}+
\left(E-M+\frac{A_2}{r}+\frac{A_1}{r}\right)G_{\ell j}&&= 0.
\end{eqnarray}
Note that the interaction in these two equations  can be
diagonalized so that we obtain

\begin{equation}
\label{A+-} A^+\hat{G}\propto\hat{F}, \quad A^-\hat{F}
\propto\hat{G}
\end{equation}
where

\begin{equation}
A^{\pm}= \pm\frac{d}{dr}+\frac{\lambda}{r}-\frac{EA_1 + MA_2}{\lambda}.
\end{equation}

These relations are similar to the relations between the two
components of the eigenfunctions of a "supersymmetric" Hamiltonian
which satisfies the following Lie graded algebra

\begin{eqnarray}
\label{Hsusy}
{\cal H}&&= [{\bf Q, Q^{\dagger}}]_+={\bf Q Q^{\dagger}}+
{\bf Q^{\dagger}}{\bf Q}, \quad [{\cal H}, {\bf Q^{\dagger}}]_-=0=
[{\cal H},{\bf Q}]_-
\end{eqnarray}
with the following representation

\begin{equation}
\label{repr}
{\bf Q}= \left(\begin{array}{cc}0 & 0\\A^- & 0
\end{array}\right), \qquad
{\cal H}= \left(\begin{array}{cc} H_-=A^+A^- & 0\\0 & H_+=A^-A^+
\end{array}\right), \quad \Phi_{SUSY}= \left(\begin{array}{cc}F\\G\end{array}\right).
\end{equation}
Note that the supercharges are nilpotent operators, viz.,
$({\bf Q^{\dagger}})^2=0= {\bf Q}^2.$

Thus, using the shape invariant Hamiltonians $H_{\pm}$ we obtain the
energy eigenvalues associated to the component $\hat{F}^n$ given by

\begin{equation}
E_{n}= \sqrt{\frac{M^2}{1+\frac{\gamma^2_n}{(\sqrt{k^2 -
\gamma^2_n}+n)^2}}} \quad n= 0, 1, 2,\cdots, \quad
\gamma_n(E)=A_1+\frac{MA_2}{E_n}.
\end{equation}
In conclusion, we  obtain the complete set of the energy eigenvalues
of the Dirac equation for a potential which is the sum of the
Coulomb potential with a Lorentz scalar potential inversely
proportional to $r$ via shape invariance property as applied in
\cite{R04}. One of us (RLR) will make elsewhere a detailed
analysis for this problem as applied to the relativistic Coulomb
potential via SUSY shape-invariant potentials \cite{R04}.


\centerline{\bf Acknowledgments}

RLR was supported in part by CNPq (Brazilian Research Agency). He
also thanks the staff of the CBPF and CES-UFCG.

\end{document}